\documentclass[a4paper,10pt]{article}
\usepackage[utf8]{inputenc}
\usepackage{amsmath}
\usepackage[pdftex]{hyperref}
\usepackage[T1]{fontenc}
\usepackage{lmodern}					
\usepackage{amsbsy}
\usepackage{amssymb}
\usepackage{setspace}
\usepackage{wasysym}
\usepackage{graphicx}
\usepackage{geometry}
\usepackage{mathtools}
\usepackage{titling}
\usepackage{float}
\usepackage{appendix}

\usepackage{authblk}
\usepackage{colortbl}

\makeatletter
\@ifundefined{showcaptionsetup}{}{%
 \PassOptionsToPackage{caption=false}{subfig}}
\usepackage{subfig}

\title{Studying viral populations with tools from quantum spin chains}

\begin{document}

\author{Saumya Shivam$^1$, Christopher L. Baldwin$^{2,3}$, John Barton$^4$, Mehran Kardar$^5$, and  S. L. Sondhi$^1$}
\date{%
    \small{$^1$\textit{Department of Physics, Princeton University, Princeton, New Jersey 08544, USA}\\%
    $^2$\textit{National Institute of Standards and Technology, Gaithersburg, MD 20899, USA}\\
    $^3$\textit{Joint Quantum Institute, University of Maryland, College Park, MD 20742, USA}\\
    $^4$\textit{Department of Physics and Astronomy, University of California, Riverside, CA 92521, USA}\\
    $^5$\textit{Department of Physics, Massachusetts Institute of Technology, Cambridge, Massachusetts 02139, USA}}\\[2ex]%
}
\maketitle




\begin{abstract}
We study Eigen's model of quasi-species ~\cite{Eigen1971Selforganization}, characterized by sequences
that replicate with a specified fitness and mutate independently at single sites.
The evolution of the population vector in time is then closely related to that of
quantum spins in imaginary time. 
We employ multiple perspectives and tools from interacting quantum systems to examine growth and collapse of realistic viral populations, specifically certain HIV proteins.
All approaches used, including the simplest perturbation theory, give consistent results.
\end{abstract}

\section{Introduction}\label{intro}

A central concept in population genetics is an idealized fitness $F$, which (in the absence of competition,
 resource limitation, etc.) governs the exponential growth in the number of individuals $N$
 according to
$\frac{dN}{dt}=FN$~\cite{Haldane1927Mathematical}.
However, mutations diversify the genetic make-up (genotype) of the population, and modify its overall 
fitness.
Eigen introduced the concept of {\bf quasi-species} to describe the `cloud' of closely
related genotypes. 
Eigen's model considers a population composed of a set of sub-populations (labelled $\{a\}$), 
each contributing $N_a$ individuals that coexist in a  rapidly evolving larger community of (varying) total size $N=\sum_aN_a$~\cite{Eigen1971Selforganization}.
Mutations between quasi-species further diversify the composition of the population.
If the sub-populations $\{a\}$ are assigned fitness values $\{F_a\}$, 
and mutations from sub-population $a$ to $b$ occur at rates $W_{ba}$, 
the makeup of the population changes over time according to
\begin{equation}\label{Eigen}
\frac{dN_a}{dt}=F_aN_a+\sum_{b\neq a}[W_{ab}N_b-W_{ba}N_a]\,.
\end{equation}
In terms of the fractions $x_a=N_a/N$,
the total population size $N=\sum_aN_a$ grows as
$\frac{dN}{dt}=\overline F N$.
The \textit{mean fitness} $\overline F = \sum_a F_a x_a$, must be non-negative
for the quasi-population to survive at long times.

At the molecular level, genetic information is maintained in the sequence of bases
of nucleotides (e.g. in the sequence of the DNA or RNA of a virus).
For modeling purposes, we assume that this information appears as a sequence of $L \gg 1$ characters
(e.g. one of 4 nucleotides of DNA, or 20 amino acids of a protein).
For simplicity of presentation (as well as practicality, as in the case of HIV proteins described below)
we adapt a binary representation, $n_i=(0,1)$ or $\sigma^z_i=(-1,+1)$, for $i=1,2,\cdots, L$, 
to indicate whether the site $i$ is in its wild-type (consensus) or a mutated state.
In this representation of a genotype, its fitness is some function of its sequence
$F_a=F[\left\{\sigma_i^z\right\}]\,$. Note that by our definition of a consensus sequence, it is possible for a sequence to have a higher fitness than the consensus sequence, 
since the consensus is based only on single-site  frequencies.

In a simple model of mutations, the state of each site (independently of other sites) 
changes away from consensus at the rate $\mu_f$, and reverts back to wild-type at the rate $\mu_b$. 
Equation~(\ref{Eigen}) is linear, and its right hand side can be regarded as representing the action
of a matrix $H$ on a (column) vector ${\bf N}$ containing the sub-population sizes $\{N_a\}$.
For $\mu_f=\mu_b=\mu$, the action of mutations on site $i$ can be represented by
the Pauli matrix $\mu \sigma_i^x$, such that
\begin{equation}\label{Hdynamics}
\frac{d{\bf N}}{dt} = -H {\bf N}\,,\quad{\rm with}\quad
H = -F[\left\{\sigma_i^z\right\}] - \mu\sum_{i=1}^L (\sigma_i^x-{\bf 1})\,.
\end{equation}
The above equation provides an analogy to the imaginary-time evolution of spins in a quantum chain,
governed by the Hamiltonian $H$.
The fitness function $F[\left\{\sigma_i^z\right\}]$ corresponds to interactions between spins,
while mutations are implemented through the transverse magnetic field $\mu$. 

The analogy between Eigen's quasi-species model and quantum chains has been noted and
explored in a number of references~\cite{Wagner1998Ising,Baake2001Mutation,baake_ising_1997,baake_quantum_1998,hermisson2001four,saakian_eigen_2004}. 
The behaviour of genetic structure of a quasi-species model and magnetization of the corresponding quantum spin model for some idealized fitness landscapes (like an Ising chain or a mean field Hamiltonian) have been studied in Refs.~\cite{Wagner1998Ising,Baake2001Mutation,baake_ising_1997}. Hermisson et al.~\cite{hermisson2001four} generalize the binary representation of DNA sequences to the four nucleotides, leading to a corresponding four state quantum chain. Saakian \& Hu~\cite{saakian_eigen_2004} consider mutations between sequences separated by more than one Hamming distance 
(see also Ref.~\cite{Rumschitzki1987Spectral}).
As is well known in physics literature, a quantum system in $d$ dimensions is equivalent to a classical system in $d+1$ dimensions with discretized time. Leuth\"ausser~\cite{LeuthAusser1987Statistical} explored a similar analogy between the Eigen model and a two dimensional Ising system. 

There are, however, important differences between  quasi-species evolution,
and the dynamics of a quantum chain (in real time):
\par\noindent$\bullet$ While the Hamiltonian $H$ is real and  symmetric, the 
evolution of ${\bf N}$ is not unitary, in that the overall 
population size $N(t)=\sum_aN_a(t)$ changes as a function of time.
By contrast, time variation according to $\frac{d{\bf N}}{dt} = - iH{\bf N}$ preserves the
norm of the vector ${\bf N}$, such that $\{|N_a|^2\}$ can be regarded as probabilities.
The natural set of probabilities for the evolving population are the proportions $\{x_a(t)=N_a(t)/N(t)\}$.
\par\noindent$\bullet$ Even the symmetry of $H$ is an artifact of the
simplification $\mu_f=\mu_b=\mu$.  In the biological context, it is more likely that forward
and mutation rates are not the same, leading to the replacement of 
$\mu( \sigma_i^x-{\bf 1})$ for $\mu_f\neq\mu_b$ with the asymmetric matrix  
 $\mu_f\sigma^++\mu_b\sigma^-$~\cite{saakian_eigen_2004}.
Despite these differences, there are aspects of the dynamics that are common to the two systems, in certain special cases. For quantum systems at low temperature, one is often interested in the low energy properties, which are determined by the eigenvectors of $H$ corresponding to the lowest eigenvalues. Similarly, the long-time behavior in the quasi-species evolution is governed by the few largest
eigenvectors of the matrix $-H$, i.e., the exact same eigenstates. 
In particular, the ground state energy is (minus) the mean fitness, while
the ground state vector characterizes the prevalence of mutations in the quasi-species population. 
Together, they determine the eventual fate of the quasi-species as described below:

\par\noindent$\bullet$ {\bf Error threshold:}
Eigen introduced the concept of an error catastrophe~\cite{Eigen1971Selforganization}  by considering a fitness function
with a single peak on a uniform background (for one or more mutations), which can be described by
\begin{equation}\label{EigenF}
F_{\rm Eigen}=F_\mu+(F_0-F_\mu)\prod_{i=1}^L\frac{1-\sigma_i^z}{2} \,.
\end{equation}
In words, the wild type has fitness $F_0$ while any mutation reduces the fitness to $F_\mu<F_0$.
Upon increasing the mutation rate $\mu$, there is a transition when the fraction
of population in the fit state decreases dramatically.
This is a genuine singularity (phase transition) in the limit $L \to \infty$, although the threshold $\mu_c$ scales as $1/L$.
More precisely, one can show that for $\mu < \mu_c = (F_0 - F_{\mu}) / L$, the fraction $x_\ell$ of the population having $\ell$ mutations falls off exponentially as
\begin{equation} \label{eq:Eigen_trapped_state}
x_\ell \sim \left( \frac{\mu}{\mu_c} \right) ^\ell,
\end{equation}
whereas for $\mu > \mu_c$, nearly the entire population has roughly $L/2$ mutations.

The fitness function in Eigen's case is of course highly artificial, and questions remain as to whether the concept 
of error threshold is applicable to more realistic landscapes (see also Ref.~\cite{Galluccio1997Exact}).
Nonetheless, sharp transitions like this, in which the relevant quantity is zero on one side and non-zero on the other, are in fact quite common in the context of quantum spin systems.
Examples include the transition between ferromagnetic and paramagnetic equilibrium phases~\cite{Sachdev2011,Suzuki2013}, frozen versus thermalizing spin glasses~\cite{Jorg2008Simple,Bapst2013Quantum,Baldwin2017Clustering}, and many-body localization-delocalization transitions~\cite{Pal2010Many,Nandkishore2015Many,Abanin2019Colloquium}.

\par\noindent$\bullet$ {\bf Fatal mutation load:}
In Eigen's model, the quasi-species population can still grow if $F_\mu>0$, as each sequence
is viable.
However, the population can also disappear upon increased mutation because of
the reduction in overall fitness.
As an example, consider the so-called Mount Fuji Landscape~\cite{baake_ising_1997}:
\begin{equation}\label{MFL}
F_{\rm MFL}=-LF_0-h\sum_{i=1}^L \sigma_i^z ,
\end{equation}
specifically with $0 < F_0 < h$
The most fit sequence has a fitness $L(h-F_0)$, with any mutation (independently) carrying a load of $2h$. 
With increasing mutation rate, the quasi-species cloud acquires more mutations, but 
(unlike the case of error threshold) remains anchored to the wild-type state. 
Since the problem is equivalent to independent quantum spins, it is easy to identify
the ground state as corresponding to all spins tilted along $(-h,\mu)$, leading to the
mean fitness of $\bar F(\mu)=L(\sqrt{h^2+\mu^2}-\mu-F_0)$.
For any mutation rate $\mu$ larger than $\mu_c = (h^2 - F_0^2)/2F_0$, $\overline{F}(\mu)$ is negative and the population will die off, despite the lack of any error threshold.

\section{Viral prevalence landscape}\label{sec2}
Several other forms of fitness landscape have been proposed and explored in the literature.
We would like to explore the quantum analogy, identifying mechanisms
that lead to population collapse (e.g. by error catastrophe or high mutation load).
To this end, we will focus on a particular class of landscapes constructed from
the prevalence of observed sequences for a virus.

The massive size of the space of possible protein sequences makes it impossible to estimate virus prevalence simply by counting sequences. Each site in a protein sequence can be one of 20 amino acids, and typical protein sequences have lengths on the order of hundreds of sites. However, the available data typically consists of only a few thousand sequences. 

A common approach to this problem is to search for the simplest probability distribution (by which we mean the one with the largest entropy) over the space of protein sequences that is capable of reproducing the single site and pairwise frequencies of amino acids in the data~\cite{cocco2018inverse}. To further reduce complexity, one can represent protein sequences with a reduced or even binary (zero for `wild-type' and one for `mutant') amino acid alphabet~\cite{ferguson2013translating,barton2016ace,rizzato2020inference}. In the binary case, the maximum entropy model capable of reproducing the empirical correlations is an Ising model with local fields $h_i$ and pairwise couplings $J_{ij}$. Finding the appropriate fields and couplings to reproduce the correlations in the data is a challenging statistical problem~\cite{cocco2018inverse}. For the models presented here, we applied the adaptive cluster expansion method~\cite{cocco2011adaptive,barton2016ace} to infer field and coupling parameters that reproduced mutant frequencies and correlations observed in sequence data from a variety of HIV proteins~\cite{Barton2015Scaling}.

Following the assumption that the most prevalent viral sequences are likely to also have the highest fitnesses, the prevalence landscape can be used as a proxy for fitness. The simple connection between prevalence and fitness could be obscured due to strong interactions between HIV and the immune system, which drives the virus to accumulate mutations rapidly. Careful modeling suggests, however, that prevalence is a good proxy for fitness for viruses such as HIV, which produces chronic infections and stimulates a diverse range of immune responses~\cite{shekhar2013spin}. This assumption is also well-supported by experiments that tested the effects of different mutations on HIV replication~\cite{ferguson2013translating,mann2014fitness,louie2018fitness}.

\section{Estimating mean fitness}\label{sec3}

We study population collapse for a few HIV proteins for which the fitness 
has been estimated to have the form~\cite{Barton2015Scaling,Chen2019MeanField}
\begin{equation}\label{IsingF}
    F[\{\sigma^z_i\}]=F_0-\sum_ih_i\sigma^z_i-\sum_{i<j}J_{ij}\sigma^z_i\sigma^z_j\,,
\end{equation}
where $\sigma_i$ are the Pauli matrices and $F_0$ is a constant. 
Including the effect of mutations as described in Eq.~\ref{Hdynamics} leads to a dynamics governed by the Hamiltonian
\begin{equation} \label{eq:effective_Hamiltonian}
\begin{aligned}
H =& \; -F[\{\sigma_i^z\}] - \mu\sum_i(\sigma^x_i-\bf{1}) \\
=& \; -F_0+\sum_ih_i\sigma^z_i+\sum_{i<j}J_{ij}\sigma^z_i\sigma^z_j-\mu\sum_i(\sigma^x_i-\bf{1}).
\end{aligned}
\end{equation}
We apply to this Hamiltonian three simple techniques, all common for studying quantum spin systems: first-order perturbation theory in $\mu$, second-order perturbation theory, and a (non-perturbative) mean-field approximation.
As will be shown, all three give consistent results, and whatever discrepancies exist can be easily understood.\\

\subsection{First-order perturbation theory}

While sophisticated methods from quantum spin chains can certainly be applied to such a system, in practice we find that standard perturbation theory provides a more than adequate tool for identifying population collapse.

Let us denote the energy of the consensus state at $\mu = 0$ by $E_0$.
Recall that the corresponding state is defined to have all $\sigma_i^z = -1$.
To first order in perturbation theory, the change in energy is the expectation value
of $\mu\sum_i(\sigma^x_i-\bf{1})$ in this state. 
The expectation value of $\sigma_i^x$ being zero in any $\sigma_i^z$ eigenstate, the perturbed energy is simply
\begin{align}
    E = E_0 + \mu L\,,
\end{align}
which becomes positive at the threshold value (remember that $E_0$ must be negative)
\begin{align}\label{1storder}
    \mu_c=-\frac{E_0}{L}.
\end{align}

\subsection{Second-order perturbation theory}

The second-order corrections to the energy, assuming non-degenerate levels, give an expression
\begin{align} \label{eq:second_order_expression}
E = E_0 + \mu L + \mu^2\sum_{i}\frac{1}{E_0-E_i}\,,
\end{align}
where the sum is over configurations with one spin flipped. The threshold can then be found by solving the quadratic equation
\begin{align}\label{2ndorder}
E_0 + \mu_c L + \mu_c^2\sum_{i}\frac{1}{E_0-E_i} = 0.
\end{align}

Note from Eq.~\eqref{eq:second_order_expression} that if $E_0$ is less than all $E_i$ (as is often the case, and must be if the consensus is the true ground state), then the second-order terms are necessarily negative.
The energy is lower than the first-order estimate, and thus the threshold mutation rate is larger.

\subsection{Mean-field approximation}
As an alternative to the perturbative calculations, we also employ a mean field approach to directly find an approximation to the ground state of the quantum Hamiltonian.
In this procedure the interacting Hamiltonian 
\begin{align}
    H=-F_0+\sum_ih_i\sigma^z_i+\sum_{i<j}J_{ij}\sigma^z_i\sigma^z_j-\mu\sum_i(\sigma^x_i-1)\,,
\end{align}
is approximated by that describing non-interacting spins, as
\begin{align}
    H_{MF}=-F_0+\sum_ih^{eff}_i\sigma^z_i-\mu\sum_i(\sigma^x_i-1)\,.
\end{align}
The effective field at each site is expressed in terms of the average $z$ component of spins at connected sites as 
\begin{align}
    h^{eff}_i=h_i+\sum_{j\neq i}J_{ij}\langle \sigma^z_j \rangle\,\quad{\rm where}\quad \langle \sigma^z_j \rangle=-\frac{h^{eff}_i}{\sqrt{(h^{eff}_i)^2+\mu^2}}\,.
\end{align}

To solve the set of coupled equations for $\{h^{eff}_i\}$, we use an iterative procedure,
starting with the $\{h_i\}$, alternately computing the expectation values of spins and the corresponding effective fields
until they  converge.
The mean-fitness is finally computed as
\begin{equation}\label{Fmf}
\bar F_{MF}(\mu)=F_0+\sum_{i=1}^N\left(\sqrt{(h^{eff}_i)^2+\mu^2}-\mu\right)\,,
\end{equation}
from which we obtain the mutation threshold by setting $\bar F_{MF}(\mu) = 0$.

\subsection{Results on mutation threshold for HIV proteins}

We considered five HIV proteins of variable lengths:
p24 ($L=231$), gag ($L=490$), integrase ($L=286$), nef ($L=206$), and protease ($L=99$). 
The Ising on-site and exchange fields for each protein were estimated previously~\cite{Barton2015Scaling,Chen2019MeanField}
from prevalence of different sequences in collected data.
We then multiplied the prevalence landscape by a factor of $\beta=0.023$/day, and added an overall constant  of $F_0=1.6$/day 
to construct a fitness landscape. These choices were dictated by typical viral loads in patients, as discussed in  Ref.~\cite{Chen2019MeanField}.\\

\begin{figure}
    \centering
    \includegraphics[width=12cm]{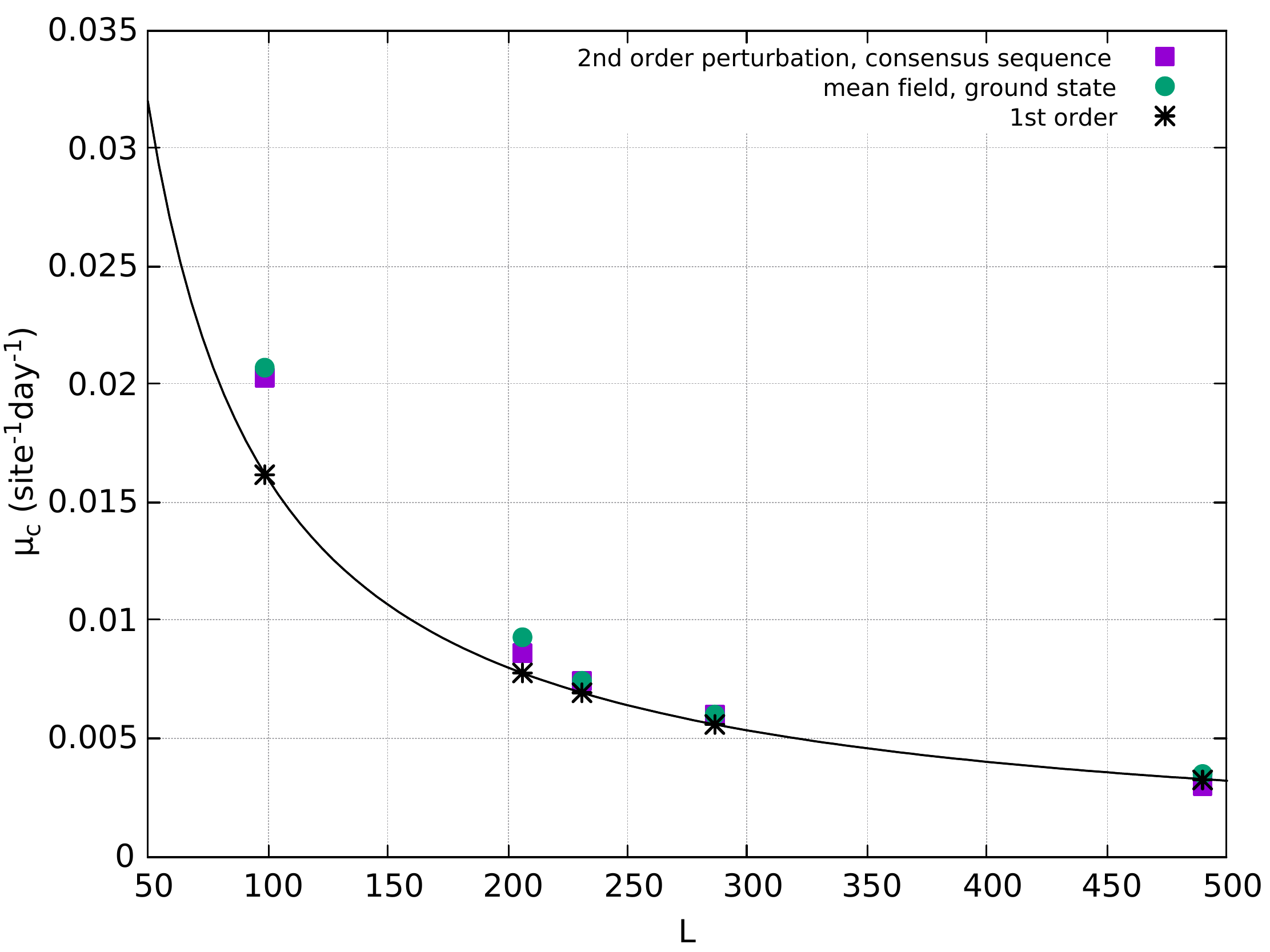}
    \caption{Threshold mutation rates from mean-field approximation, compared to those from perturbing around the consensus sequence
    for five HIV proteins gag ($L=490$), integrase ($L=286$), nef ($L=206$), p24 ($L=231$), and protease ($L=99$). The solid line is Eq.~\eqref{1storder} assuming constant $E_0$.}
    \label{fig:critical_mu}
\end{figure}

In all cases, we both performed perturbation theory around the consensus state and applied the mean-field approximation.
It will be important to note that the consensus is not the fittest state for the gag, nef, and protease proteins, whereas it is the fittest for p24 and integrase.
The resulting threshold values $\mu_c$ are shown in Fig.~\ref{fig:critical_mu}.

All estimates for $\mu_c$ are quite consistent.
There are some discrepancies, but they are all as expected.
First, it is indeed the case that the first-order estimates are typically lower than the second-order values (the gag protein is the sole exception out of those considered).
For p24 ($\mu_c=0.007$/site/day) and integrase ($\mu_c=0.006$/site/day), the consensus sequence corresponds to the ground state and thus we find quite good agreement between the second-order and mean-field results.
Yet for gag, nef, and protease, the perturbation theory values only inform us of the mutation rate beyond which the consensus and all similar sequences die out.
Alternate quasi-species having lower energies will survive for slightly larger $\mu$.
Since the mean field theory is able to (approximately) identify the true ground state, regardless of whether it agrees with the consensus or not, we expect it to predict higher values for $\mu_c$.
This is indeed what we find for the three proteins in question.
We also note that the threshold mutation rates obtained for those are higher as compared to some obtained for other HIV proteins via alternate methods ($\sim 10^{-4}$/site/day)~\cite{Gupta2015Scaling,Hart2015Error}.
\\

\subsection{Effect of mutations on comparison with experimental fitness}
\begin{figure}
    \centering
    \includegraphics[width=12cm]{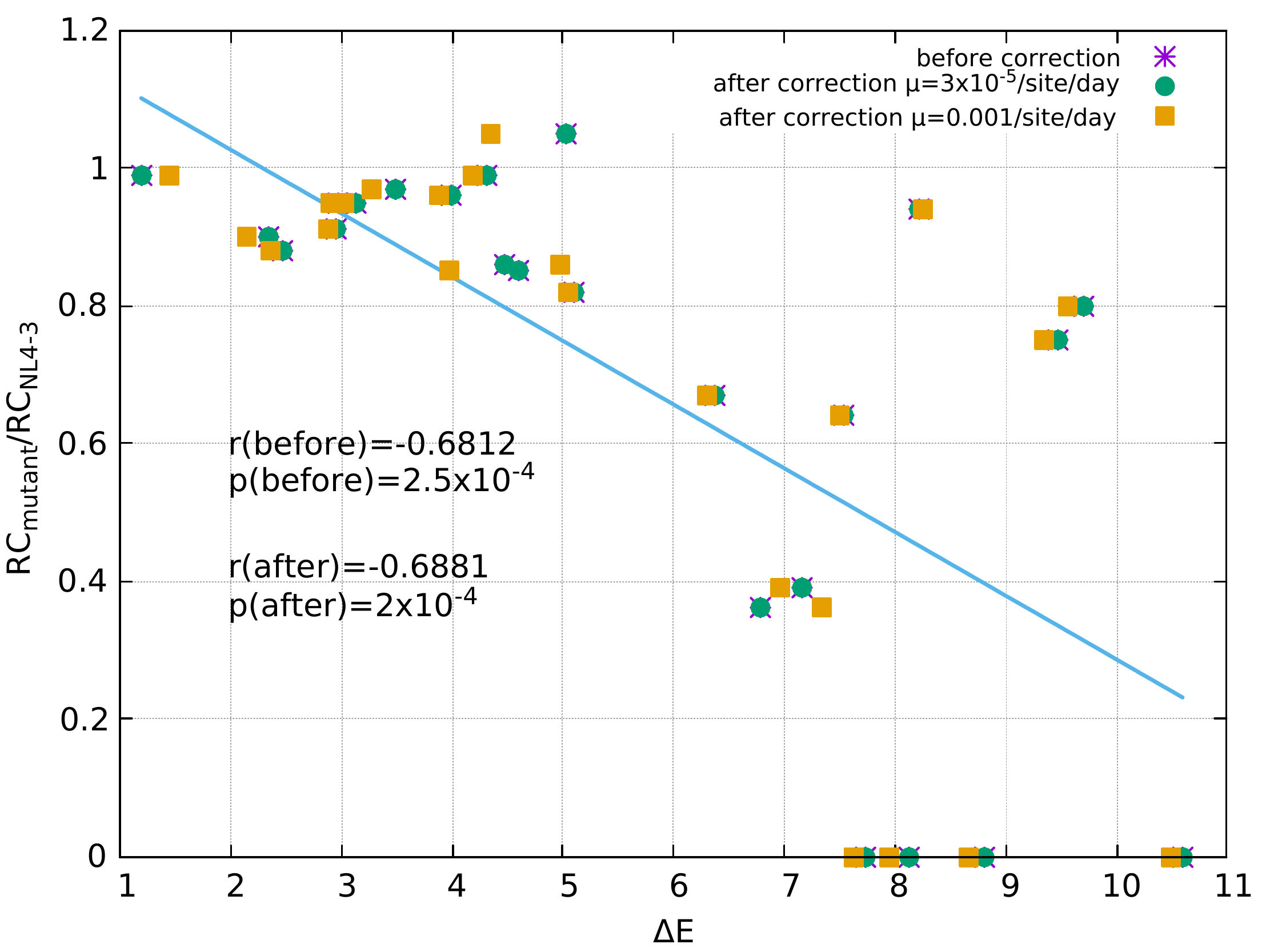}
    \caption{Relationship between the predicted energy difference of some mutants relative to the consensus sequence and their replication capacities~\cite{mann2014fitness}, for the gag protein (the NL4-3 sequence is the consensus sequence). The Pearson's correlation coefficient r=-0.6812 and p(two tailed)= $2.5\times10^{-4}$, with n=24. Correction to energies using second order perturbation theory (Eq \ref{eq:second_order_expression}) leads to negligible changes for a realistic mutation rate $\mu=3\times10^{-5}$/site/day. But considering a higher mutation rate (0.001/site/day) shows that the correction can vary both in the sign and magnitude for different sequences. The r(after) and p(after) correspond to the higher mutation rate.} 
    \label{fig:corrected_fitness}
\end{figure}

As discussed in Sec.~\ref{sec2}, the prevalence landscape is able to provide us an approximation of the fitness landscape, and is well supported by experiments. One quantity often measured by experiments is the replication capacity (RC) of a sequence, which is an empirical estimate of its growth rate and thus strongly associated with fitness~\cite{mann2014fitness}. Comparing the energy and the RC of some mutatant viruses relative to the consensus sequence (for the gag protein, where multiple measurements are available) reveals a good correlation between the two. This leads us to conclude that indeed the Ising landscape derived for the proteins are a good measure of viral fitness (Figure~\ref{fig:corrected_fitness}).\\

In comparisons with experimental growth data (via RC), it may be necessary to include the effect of mutations on the Ising landscape since the observed growth includes mutations as well~\cite{mann2014fitness}. We use our perturbation theory approach to compare the corrected energies (Eq.~\ref{eq:second_order_expression}), at a mutation rate typically expected in HIV proteins $\mu=10^{-5}$/site/day, with the RC data observed for some mutations in the gag protein in \cite{mann2014fitness}. Figure~\ref{fig:corrected_fitness} shows that the effect of mutations is negligible when comparing to experimental fitness. However, to see any generic effect of mutations, we considered a higher mutation rate of 0.001/site/day (which is unrealistic but still below $\mu_c$) and we find that the corrections can vary both in the sign and magnitude for the mutants considered here.

\section{Discussion}

Here we applied ideas from interacting quantum systems to study Eigen's quasispecies model.
Using past estimates for HIV fitness landscapes, which we verified to be in good agreement with experimental data, we computed the threshold mutation rate $\mu_c$ beyond which viral populations would be expected to decline due to insufficient fitness.
We have found that first-order perturbation theory works remarkably well in predicting $\mu_c$.
Even though we do not have an exact result against which to compare, neither higher-order terms nor a non-perturbative mean-field approximation gives significantly different results.

Note that the perturbative expression in Eq.~\eqref{1storder} is very similar to Eigen's original result~\cite{Eigen1971Selforganization}, even though the population collapse identified here is not due to an error catastrophe but rather due to negative mean fitness.

Hart and Ferguson recently investigated the possibility of an error catastrophe for HIV using estimated fitness landscapes~\cite{Hart2015Error}.
They found evidence for a phase transition to a high energy, or low fitness, state in the HIV protein p6, which forms a part of gag.
The transition occurs at a temperature larger than one, which is interpreted as a signal of a mutation rate that is higher than what is observed in nature.
However, their approach does not estimate the critical mutation rate precisely in terms of a probability of mutation per replication cycle.

In fact, the immune system has special defenses that can force viruses to undergo an ``error catastrophe.''
APOBEC proteins cause viral hypermutation, which nearly always results in the production of defective viruses~\cite{simon2015intrinsic}.
However, APOBEC proteins can be countered by the HIV protein vif, thus allowing replication of the virus to continue unchecked.
It is clear that a detailed understanding of the ``phase diagrams'' for HIV proteins will be a useful tool for combating the virus.\\

\textit{Acknowledgements}: The authors would like to thank the Galileo Galilei Institute in Florence, Italy where a part of this work was performed.
This research was performed while CLB held an NRC Research Associateship award at the National Institute of Standards and Technology. MK is supported by NSF through grant No.~DMR-1708280.

\bibliographystyle{unsrt}
\bibliography{viralMBL.bib}   
\end{document}